\documentclass[a4paper,fleqn,usenatbib]{mnras}

\usepackage{newtxtext,newtxmath}

\usepackage[T1]{fontenc}
\usepackage{ae,aecompl}

\usepackage{graphicx}	
\usepackage{amssymb}	

\def\be{\begin{equation}}
\def\ee{\end{equation}}
\def\bea{\begin{eqnarray}}
\def\eea{\end{eqnarray}}
\def\const{\mathrm{const}}
\def\md{{\rm d}}
\def\J1106{{J1106$-$3647}}
\newcommand{\mods}{}
\newcommand{\mmods}{}

\defcitealias{bignalletal2006}{Bignall et al., 2006}
\defcitealias{debruynmacquart2015}{de Bruyn \& Macquart, 2015}
\defcitealias{dennettthorpedebruyn2003}{Dennett-Thorpe  \& de Bruyn, 2003}
\defcitealias{heeschen1984}{e.g., Heeschen, 1984}
\defcitealias{pradelcharlotlestrade2006}{Pradel, Charlot \& Lestrade, 2006}
\defcitealias{rickett2011}{Rickett, 2011}
\defcitealias{rickettkedziorachudczerjauncey2002}{Rickett, Kedziora-Chudczer \& Jauncey, 2002}
\defcitealias{rickettlaziochigo2006}{Rickett, Lazio \& Chigo, 2006}
\defcitealias{tuntsovbignallwalker2013}{Tuntsov, Bignall \& Walker, 2013}
\defcitealias{vanmoorseletal1996}{van Moorsel, Kemball \& Greisen, 1996}
\defcitealias{walkerdebruynbignall2009}{Walker, de Bruyn \& Bignall, 2009}
\defcitealias{schneiderehlersfalco1992}{Schneider, Ehlers \& Falco, 1992}
\defcitealias{petterslevinewambsganss2001}{Petters, Levine \& Wambsganss, 2001}
\defcitealias{linskyrickettredfield2008}{Linsky, Rickett \& Redfield, 2008}

\title[Radio spectra of J1106-3647]{Scintillation kinks, bumps and wiggles in the radio spectrum of the quasar PMN~J1106-3647}

\author[A. V. Tuntsov et al.]{Artem V. Tuntsov,$^{1}$\thanks{E-mail: Artem.Tuntsov@manlyastrophysics.org}
Jamie Stevens,$^{2}$
Keith W. Bannister,$^{3}$
Hayley Bignall$^{4}$,
\newauthor Simon Johnston,$^{3}$
Cormac Reynolds,$^{4}$
Mark A. Walker$^{1}$\\
$^{1}$Manly Astrophysics, 15/41-42 East Esplanade, Manly 2095, Australia\\
$^{2}$CSIRO Paul Wild Observatory, 1828 Yarrie Lake Road, Narrabri, NSW 2390, Australia\\
$^{3}$CSIRO Astronomy and Space Science, PO Box 76, Epping NSW 1710, Australia\\
$^{4}$CSIRO Astronomy and Space Science, 26 Dick Perry Avenue, Kensington, WA 6151, Australia\\
}
\date{Accepted 2017 May 16. Received 2017 May 11; in original form 2017 January 31.}

\pubyear{2017}

\begin{document}
\label{firstpage}
\pagerange{\pageref{firstpage}--\pageref{lastpage}}
\maketitle

\begin{abstract}
We report radio observations of the quasar PMN~J1106-3647. Our data, taken with the Australia Telescope Compact Array, show large variations in the amplitude and shape of its spectrum, on a short time-scale. A great variety of spectral features is evident, including: sharp kinks; broad spectral peaks; and wiggles.  No two spectra are alike. We interpret the variations as interstellar scintillation of a radio source that is compact, but not point-like. Under this interpretation, complex spectral structure can arise purely refractively, under high magnification conditions, or from interference between waves that have been scattered by small-scale density fluctuations (diffractive scintillation). Both effects may be playing a role in J1106-3647, and we tentatively identify kinks with the former, and wiggles with the latter. Diffractive scintillation of AGN is uncommon, as the fringe visibility is low for all but the most compact radio sources. Refractive interpretation of the kink implies that the source has a sharp, concave boundary. Our data are consistent with a mildly boosted synchrotron source, provided the scattering material is at a distance {\mods $\sim50\,{\rm pc}$} from us. 
\end{abstract}




\section{Introduction}
The light curves of some active galactic nuclei (AGN) and quasars vary rapidly at centimetre wavelengths, a phenomenon known as intraday variability (IDV) or `flickering' \citep{heeschen1984, witzeletal1986}. Radio IDV is not intrinsic to the source, but is caused by refraction and scattering of radio waves from inhomogeneities in the ionized interstellar medium (ISM) of our Galaxy \citep{rickett2002}. In some cases the variability is very rapid, with substantial flux changes on time-scales of order an hour \citep{kedziorachudczeretal1997}, and intensive study of a few such sources has provided new insights into the small-scale structure of the ionized ISM (e.g. \citealt{dennettthorpedebruyn2003, bignalletal2003, debruynmacquart2015}. The scattering media, which we will refer to as (phase-) ``screens'', appear to be small {\mods ($\sim10^{1\pm1}\,$AU  \citealt{kedziorachudczeretal2001,rickett2011})}, dense {\mods ($n_e\sim10\,{\rm cm^{-3}}$)} \citepalias{rickett2011, tuntsovbignallwalker2013, debruynmacquart2015} and highly anisotropic \citepalias{rickettkedziorachudczerjauncey2002, dennettthorpedebruyn2003, bignalletal2006, walkerdebruynbignall2009}. 

\begin{figure*}
\includegraphics[width=0.9\linewidth]{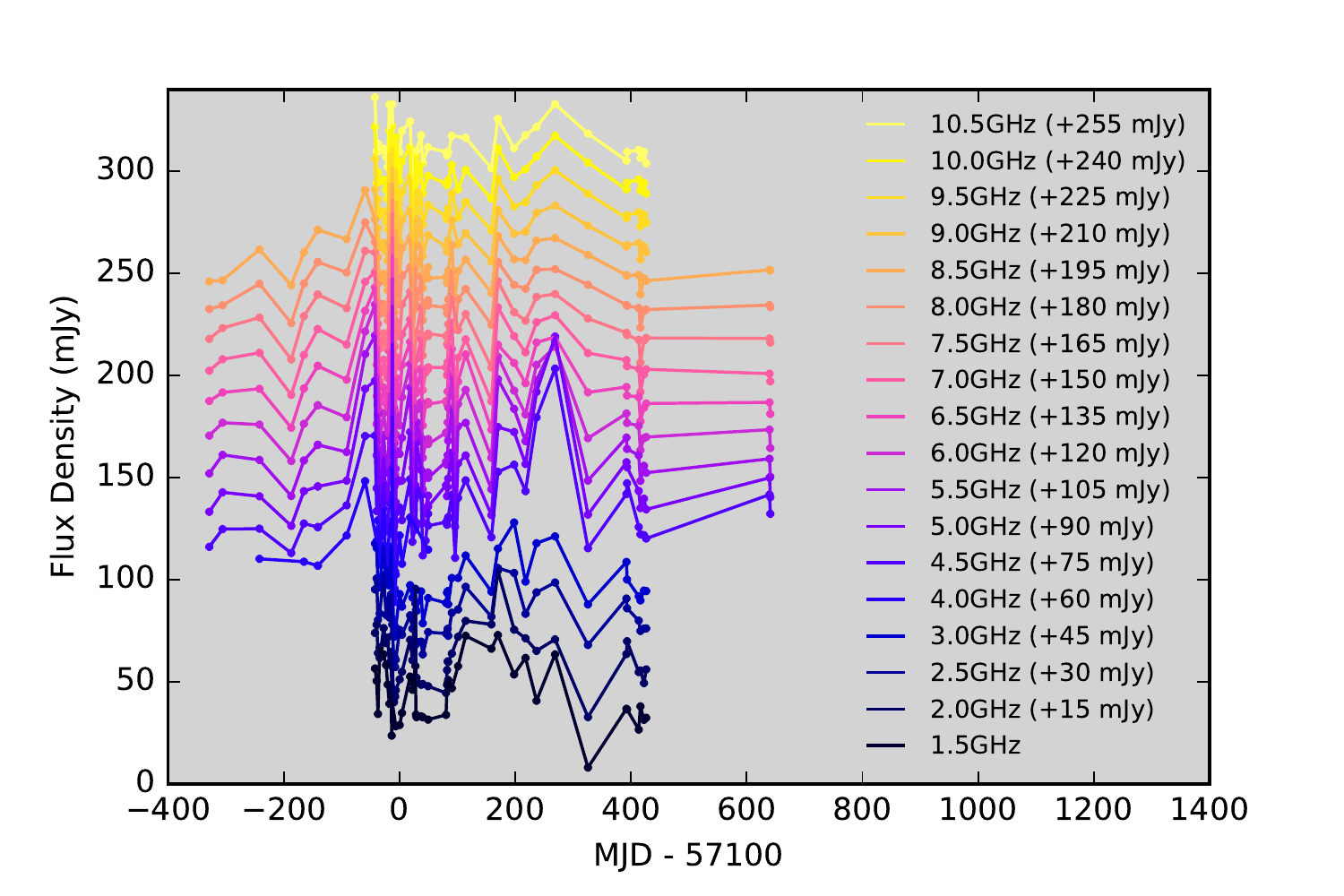}
\includegraphics[width=0.9\linewidth]{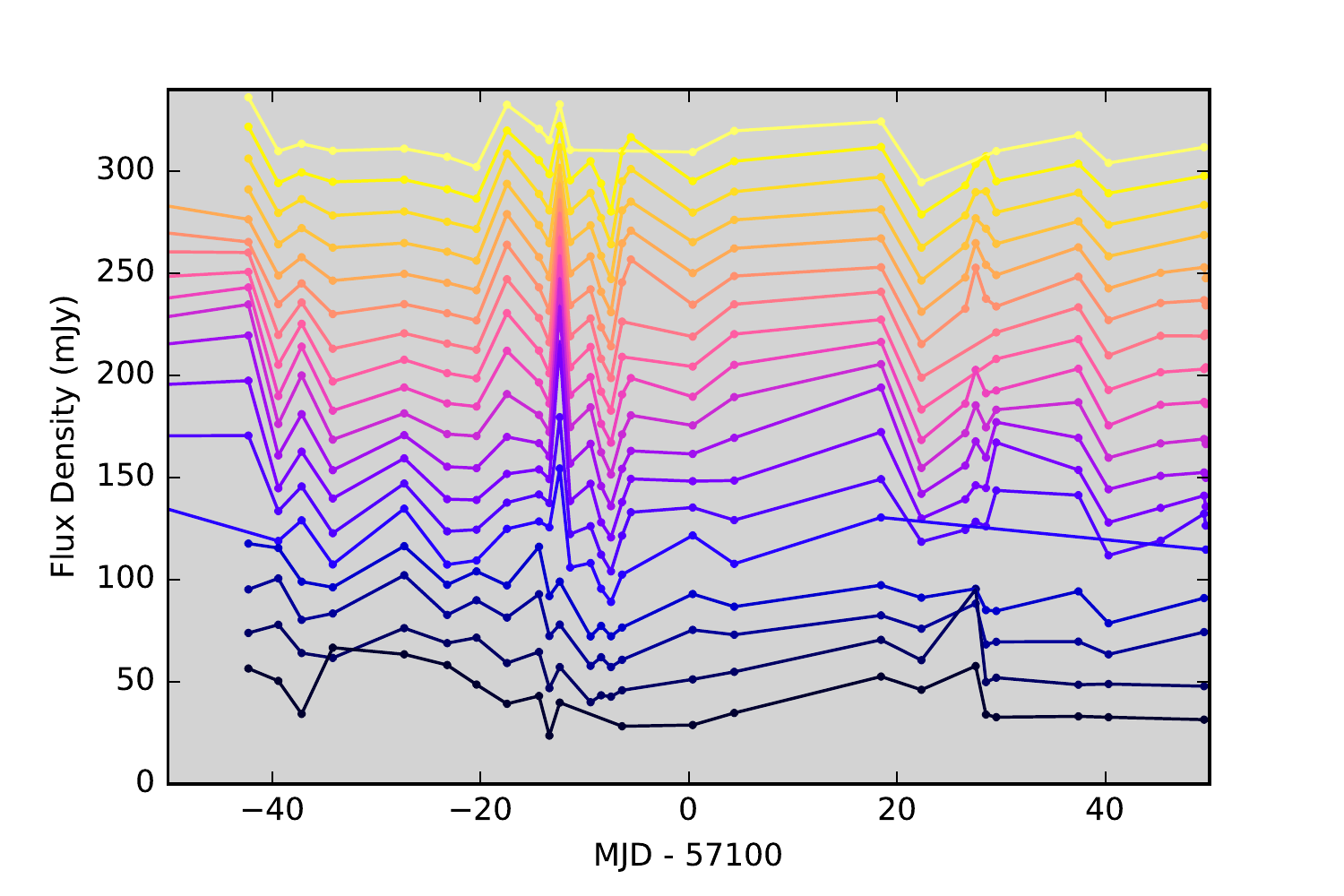}
\caption{The ATCA light curves of \J1106, colour-coded by radio-frequency, over an interval of approximately two years. Top panel: full light curves. Bottom panel: expanded view of $100\;$days around MJD~57100 (19th March 2015). Note that the amplitude of the variations is very large -- up to a factor $\sim2$ either side of 60~mJy -- and the time-scale is very short; interstellar scintillation is therefore our preferred interpretation. {\mods Light curves are displaced vertically for clarity, as annotated; the colour code is same for both panels.}}
\end{figure*}

The screens responsible for intra-hour variability are local to the Sun; possibly as close as a few parsecs \citep{dennettthorpedebruyn2000, rickettkedziorachudczerjauncey2002, bignalletal2003, debruynmacquart2015}. Nevertheless, their physical characteristics appear similar to the structures responsible for the pulsar ``parabolic arcs'' \citep{tuntsovbignallwalker2013} -- scintillation patterns formed in the dynamic spectra of pulsars \citep{stinebringetal2001, cordesetal2006,walkeretal2004} -- for which the scattering material is $\sim$kpc distant \citep{stinebringetal2001, putneystinebring2006, briskenetal2010}. Presumably, such screens pervade the Galaxy.

To date there has been no identification of any individual screen with ionized gas that is seen in direct imaging. We therefore have only one way of studying individual screens, and that is via radio wave scintillation. Moreover, the high inferred plasma pressures make it difficult to find a natural ISM context within which to construct theoretical models. {\mods Suggested physical contexts include current sheets, seen edge-on \citep{goldreichsridhar1995,penlevin2014}, boundaries of local interstellar clouds \citepalias{linskyrickettredfield2008}, and the ionised skins of tiny, molecular clouds \citep{walker2007}.} At present we are a long way from understanding the physics of the screens, and new data on the IDV phenomenon are welcome.

Here we report data on a source, PMN J1106$-$3647 \citep[\J1106, henceforth][]{wrightetal1996}, that was discovered to be rapidly variable in observations taken as part of the {\mods Australia Telescope Extreme Scattering Event (ATESE)} project (\citealt{bannisteretal2016}; 2017 in preparation) --- a survey for Extreme Scattering Events (ESEs; \citealt{fiedleretal1987, fiedleretal1994}) with the Australia Telescope Compact Array (ATCA). The ATESE project takes regular, $\sim$monthly observations, forming the survey, and more-frequent, follow-up observations of sources exhibiting anomalous spectra{\mods , i.e. spectra that are not well modeled by a smooth power law and/or have changed markedly between adjacent epochs \citep{bannisteretal2016}}. \J1106 appears in the AT20G 20~GHz radio source catalogue, with a flat radio spectrum \citep{murphyetal2010}. It is optically identified with an object of B-magnitude 19.7, but no redshift is known \citep{mahonyetal2011}. It appears to have a counterpart gamma-ray source \citep{massaroetal2015} {\mods and is considered a blazar candidate but its classification is uncertain \citep{dabruscoetal2014}}. At a Galactic latitude of $21^\circ$, it is unlikely to be a Galactic source and we assume that it is an AGN.

The temporal sampling of our data is not as good as that of some previous studies (e.g. \citealt{kedziorachudczeretal1997}), but the spectral sampling is excellent. Thanks to the ATCA's broadband receivers and correlator \citep{wilsonetal2011}, we can routinely record spectra with a resolution of 1~MHz, over an instantaneous bandwidth of $\simeq4\;$GHz. For bright {\mods ($\gtrsim 50\,\mathrm{mJy}$)} sources, such as the ATESE project targets, good signal-to-noise can be obtained in each spectral channel, in less than a minute of observing. For the most part, AGN and quasars {\mods are observed to} have quite smooth, featureless radio spectra {\mods (Bannister et al. 2017 in preparation)}. Often they can be well reproduced by spectral fits for which $\log({\rm Flux})$ is linear, or quadratic, in $\log({\rm frequency})$. By contrast {\mods to other $\sim1000$} sources in our data, \J1106 typically exhibits a great deal of spectral structure, and could not be fit by any simple analytic form. 

We anticipated that some of our ATESE targets would exhibit unusual radio spectra, as a result of refraction in the ionized ISM. The frequency-dependence of the plasma refractive index leads, in turn, to frequency-dependent magnification of the source, when a lump of plasma is in the line-of-sight, with the frequency-dependence being strongest where the magnification is highest. Indeed, the ATESE project takes advantage of this effect to pick out sources seen through strong plasma lenses. Frequency-dependent magnification, as a result of plasma lensing, may be responsible for some of the structure seen in the spectra of \J1106.

There is another way in which propagation can give rise to spectra with frequency structure: waves scattered from inhomogeneities travel slightly different paths, and constructive/destructive interference between these waves can yield maxima/minima {\mods (``fringes'')} in the measured source flux. Wave interference -- commonly referred to as ``diffractive scintillation'', in the context of interstellar radio-wave propagation -- is responsible for the large-amplitude, narrow-band structure commonly seen in radio pulsar spectra --- see \cite{rickett1990}. When diffractive scintillation is searched for in AGN, it is not usually observed \citep{dennisoncondon1981}. This is understood to be a consequence of the much larger angular size of AGN, compared to pulsars: averaging over the source intensity profile greatly reduces the fringe visibility {\mods(i.e. the contrast between maxima and minima)}. A notable counterexample is the intra-hour variable J1819+3845, for which evidence of diffractive scintillation in the 1-2~GHz band has been reported \citep{macquartdebruyn2006}.  As discussed later in this paper, diffractive scintillation might also be responsible for some of the frequency structure seen in our \J1106 spectra. 

This paper is structured as follows. In the next section we present our radio spectra of \J1106, with emphasis on the strange forms that are inexplicable as unmodified synchrotron spectra. In \S3 we offer some  interpretations of various aspects of the observed spectra. Section 4 reports VLBI observations of \J1106; and \S5 discusses some of the issues raised by our results.

\section{ATCA Observations}
The ATESE project (ATCA project code C2914) has been underway since April 2014. We survey approximately 1{,}000 compact AGN, roughly once per month. Our initial target list was selected from {\mods the combined AT20G and VLBI catalogues} \citep{murphyetal2010, petrovetal2011}, by imposing the requirements of (i) a spectral index $\alpha > -0.2$ ($S \propto \nu^\alpha$), (ii) 5~GHz flux density $> 20$~mJy, and (iii) negligible confusing flux in the ATCA primary beam at 4~GHz and above. The source \J11064 met these criteria and was therefore monitored as part of the ATESE project. We have since moved to a more Northerly sample, whose selection criteria exclude \J1106, but not before we acquired a significant dataset on \J1106.

In our monthly survey observations we record broad-band radio spectra of our sources. ATCA has two independently tuneable 2.048~GHz wide bands, each comprising 2048 1~MHz channels \citep{wilsonetal2011}. For most survey epochs the two bands are tuned to centre frequencies of 4928 and 7500 MHz respectively. We observe each source for approximately 50~s. Thermal noise is around 6~mJy in a 1 MHz channel, hence approximately 0.75~mJy when averaged to 64~MHz.

On MJD 57057 (4th Feb 2015) we began intensive monitoring of \J1106 ($\alpha={\rm11^h\,06^m\,24^s\!.04}$, $\delta={\rm-36^\circ\,46^\prime\,59^{\prime\prime}\!.2}$ J2000). In addition to a higher cadence (every few days), we increased the integration time to 120~s, yielding thermal noise of $\simeq 0.5$~mJy in 64~MHz, and broadened the observing band. We used 3 separate tunings {\mods (all done within a $\sim 15$ minute interval)} with centre frequencies of 2100, 2600, 4928, 6720, 8512 and 10304 MHz, so as to span the available frequency range of the ATCA receivers with minimal gaps between bands {\mods (the data in the $\sim100\,\mathrm{MHz}$ wide overlap centred on $\sim5.8\,\mathrm{GHz}$ were consistent between tunings and were averaged)}. During this intensive monitoring, we bracketed the observations of \J1106 with 1 minute observations of ATCA calibrator ICRF J104142.9$-$474006 {\mods (cf. lower panels in Figure~3) to check for possible instrumental systematics; no issues were identified}. {\mods This observations mode was maintained for \J1106 after we returned to $\sim$ monthly cadence in July 2015.}

We processed all data in the same way as described in \citet[supplementary]{bannisteretal2016}. Briefly: we flagged the uncalibrated data, performed standard flux-density- and bandpass-calibration, using the ATCA primary flux calibrator PKS 1934$-$638, and solved for antenna gains using self calibration on the target source. Finally, we produced vector-averaged spectra across all baselines, averaged to 64~MHz frequency resolution.

\begin{figure}
\includegraphics[width=\linewidth]{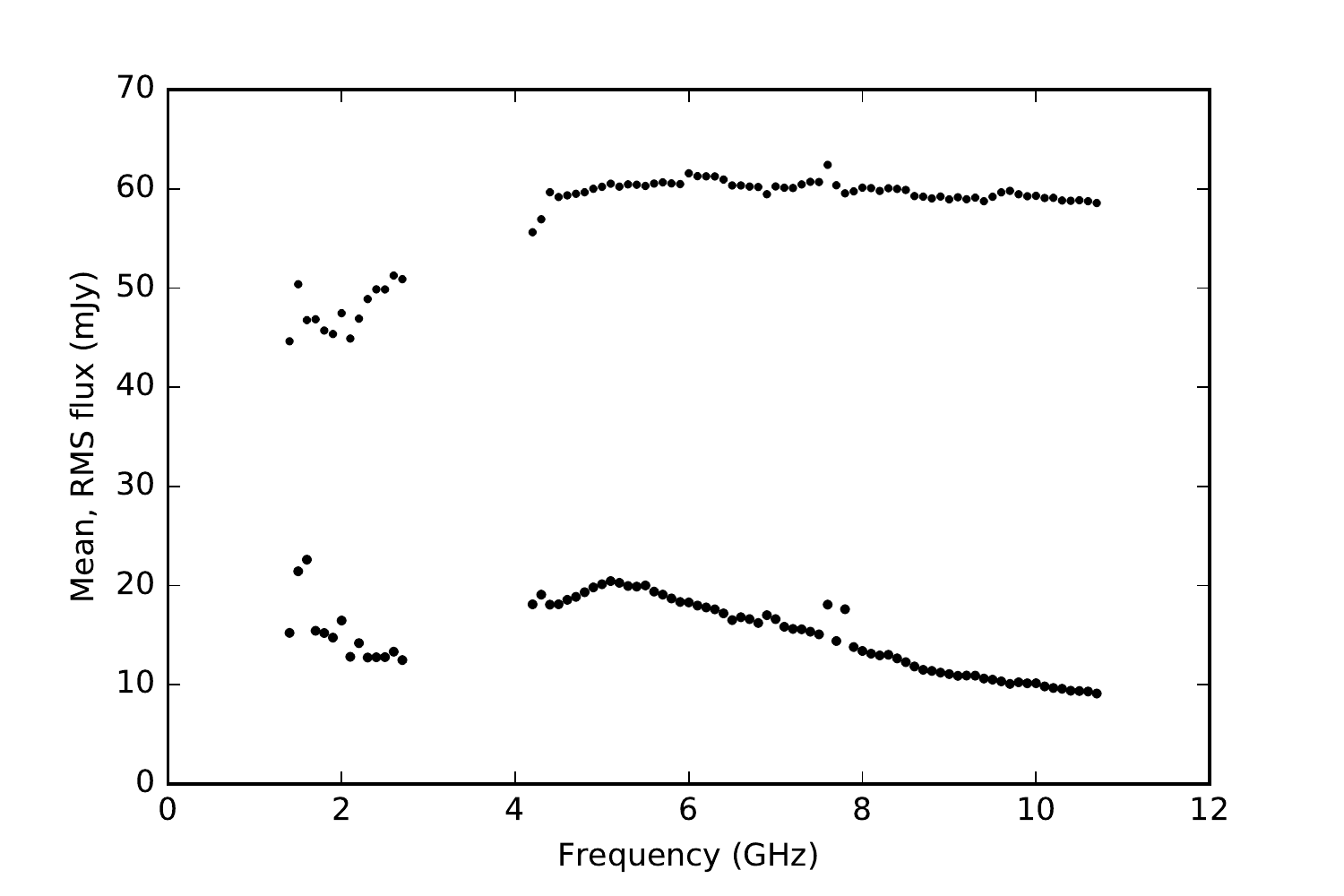}
\caption{Statistical properties of the radio flux-densities of \J1106, taken over all of the ATCA data reported in this paper. The upper set of black dots shows the mean flux-density; the lower set shows the root-mean-square flux-density variations.}
\end{figure}

Fig.~1 shows the evolution of \J1106, colour coded according to radio-frequency, over an interval of more than two years. Looking at Fig.~1, it is apparent that \J1106 varied strongly, and rapidly, for much of this time. At some frequencies the variations are as large as a factor $\sim2$ in flux-density, either side of 60~mJy. Our monthly survey observations did not adequately sample the temporal variations. Nor did the more-frequent follow-up observations, which commenced on MJD~57057 --- as can be seen in the bottom panel of Fig.~1(covering an interval of only $100\,$days). Indeed, we have instances where successive observations of \J1106, separated by only one day, exhibit completely different amplitude and shape between the two spectra.
Given these abrupt changes, we searched for short-term variability within a relatively long observing window (4~minutes long, on MJD57087.6, discussed in more detail in Section 3.3), on time-scales down to 10~seconds. No significant variation was detected. 
We note, however, that no significant intra-observation changes would be expected, if the spectrum evolved linearly in time between adjacent epochs.

Although Fig.~1 shows the \J1106 spectral variations at a coarse level, there is a great deal more information available in our spectra. Interested readers are welcome to download the reduced data on this source, or browse through the ATESE spectra of any of the $\sim$1{,}000 other sources in our sample --- instructions on how to do so, and a description of the public archive, will be given in Bannister et al (2017, in preparation). In the present paper we limit ourselves to a statistical summary of the variations, supplemented by a small number of example spectra, and we give possible interpretations for the observed spectral shapes. In broad-brush, the physical interpretation of the variations of \J1106 is clear: it is interstellar scintillation. The large amplitude and short time-scale of the observed variations preclude any other explanation for an AGN. 

\begin{figure*}
\includegraphics[width=0.48\linewidth]{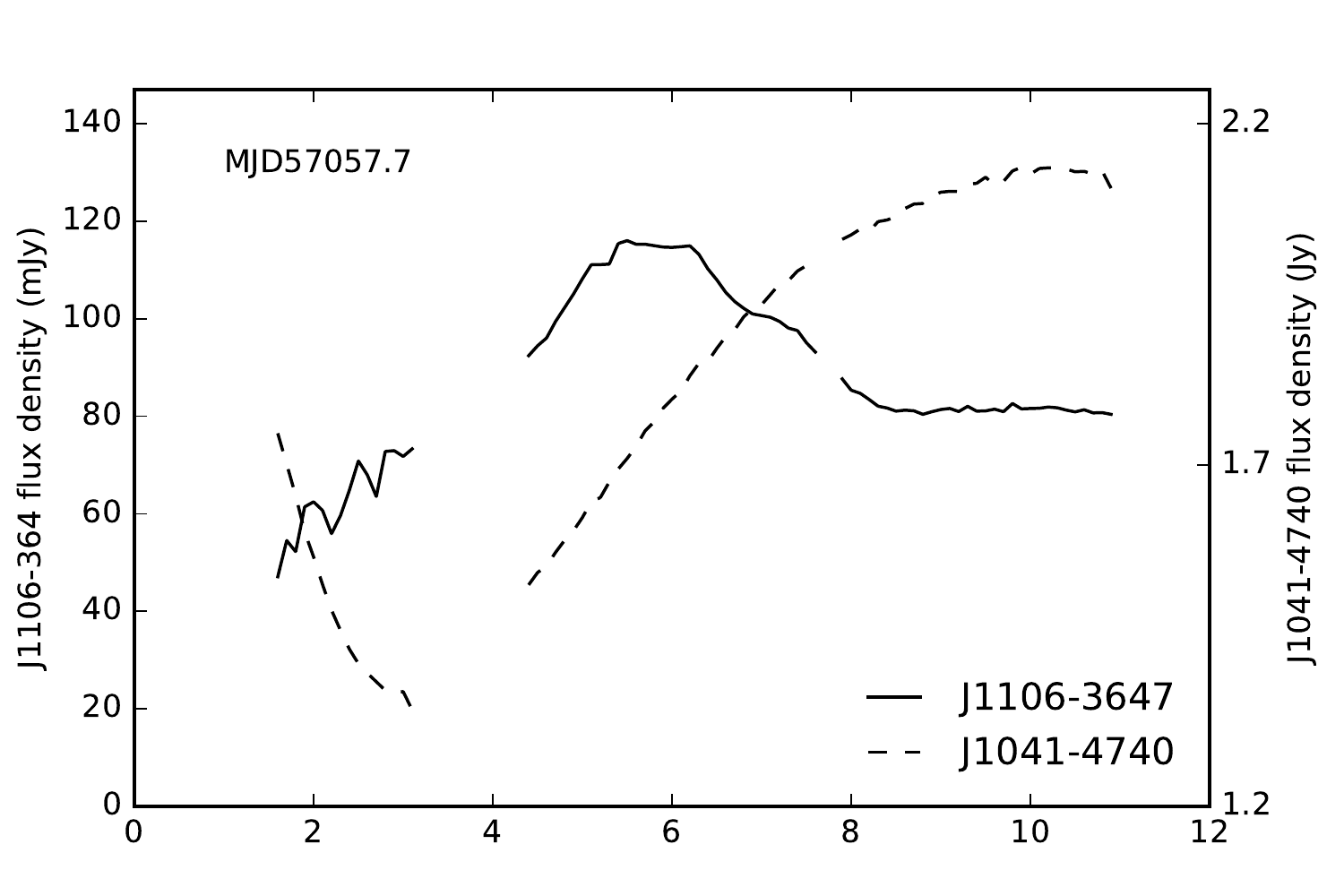}
\includegraphics[width=0.48\linewidth]{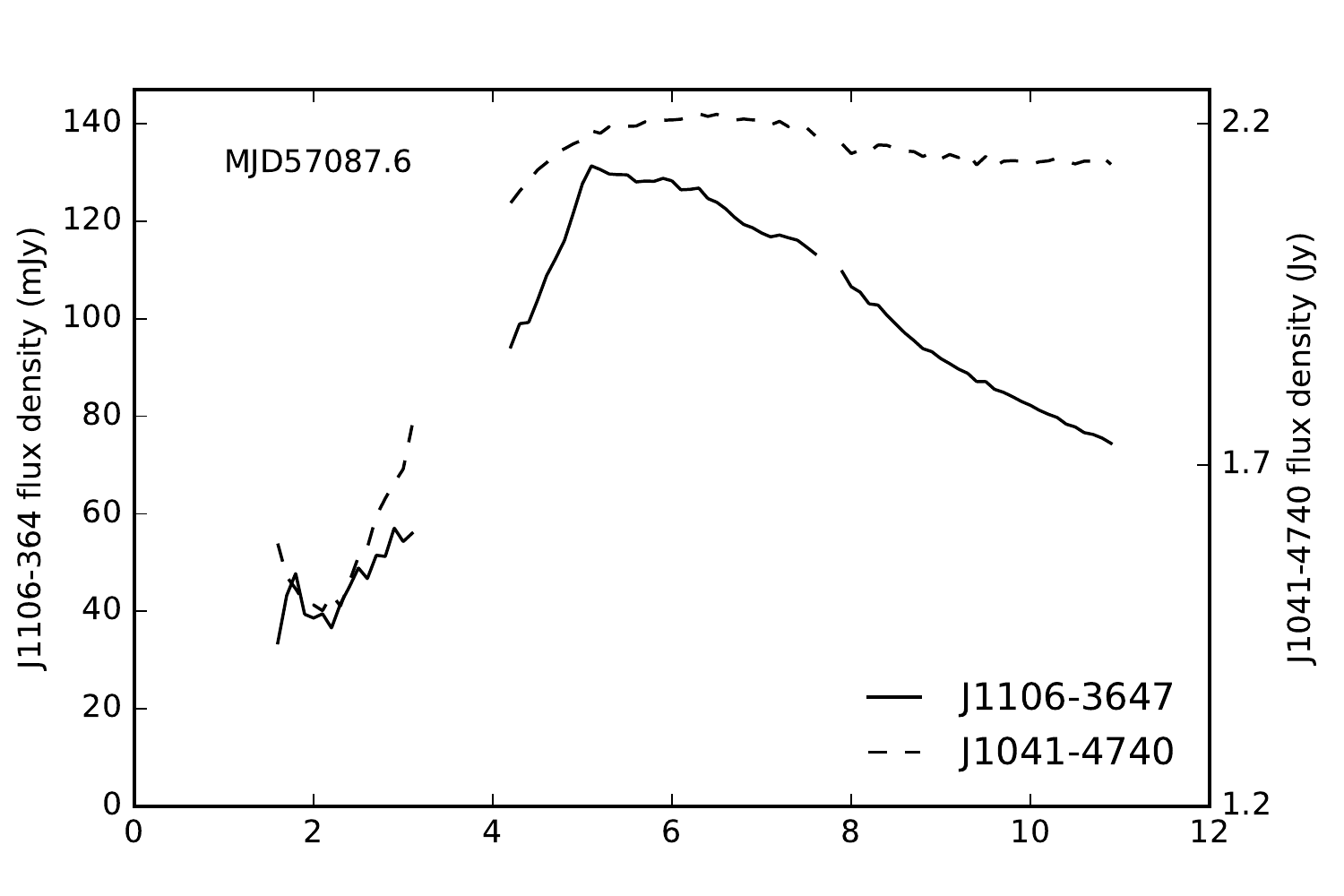}\\
\includegraphics[width=0.48\linewidth]{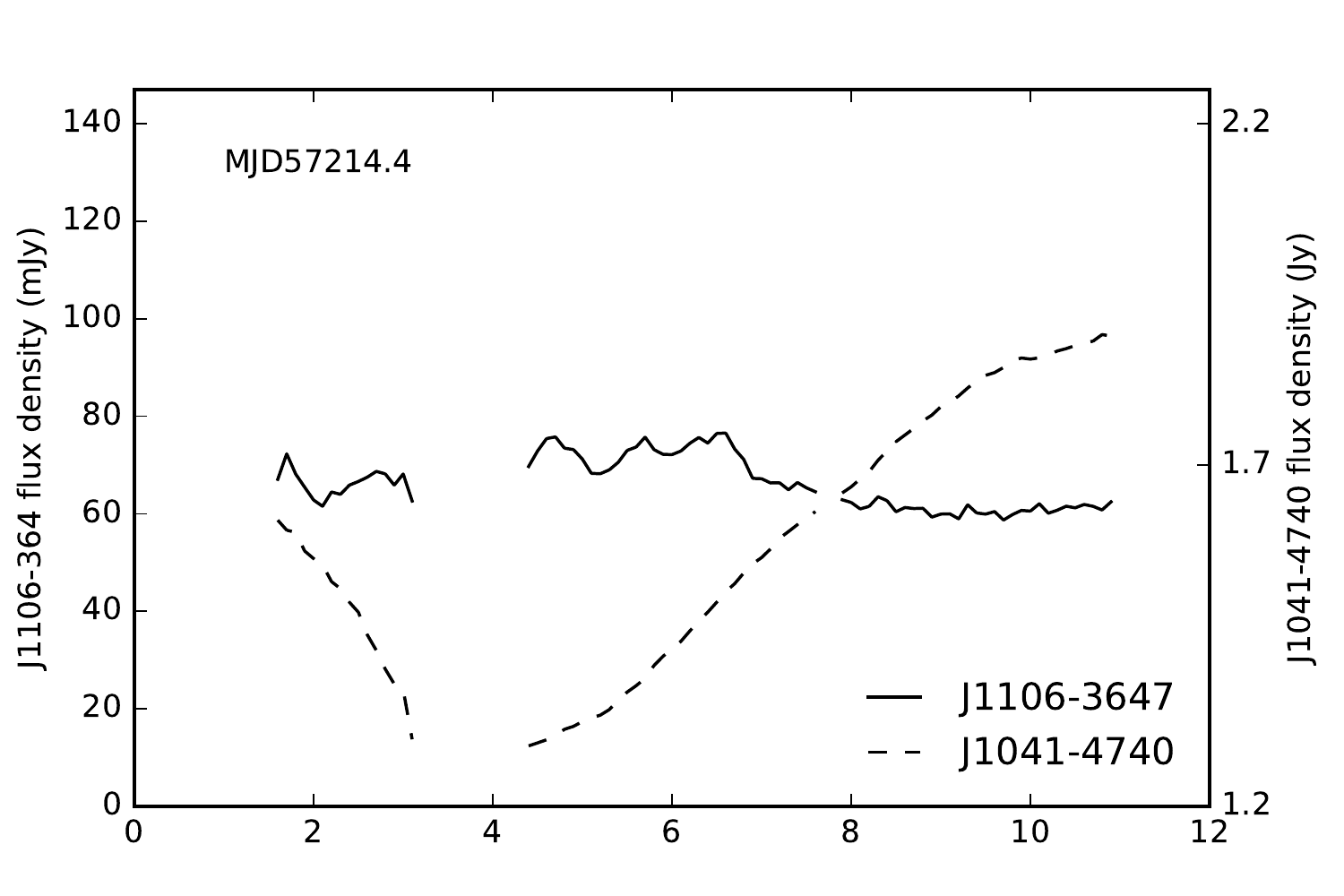}
\includegraphics[width=0.48\linewidth]{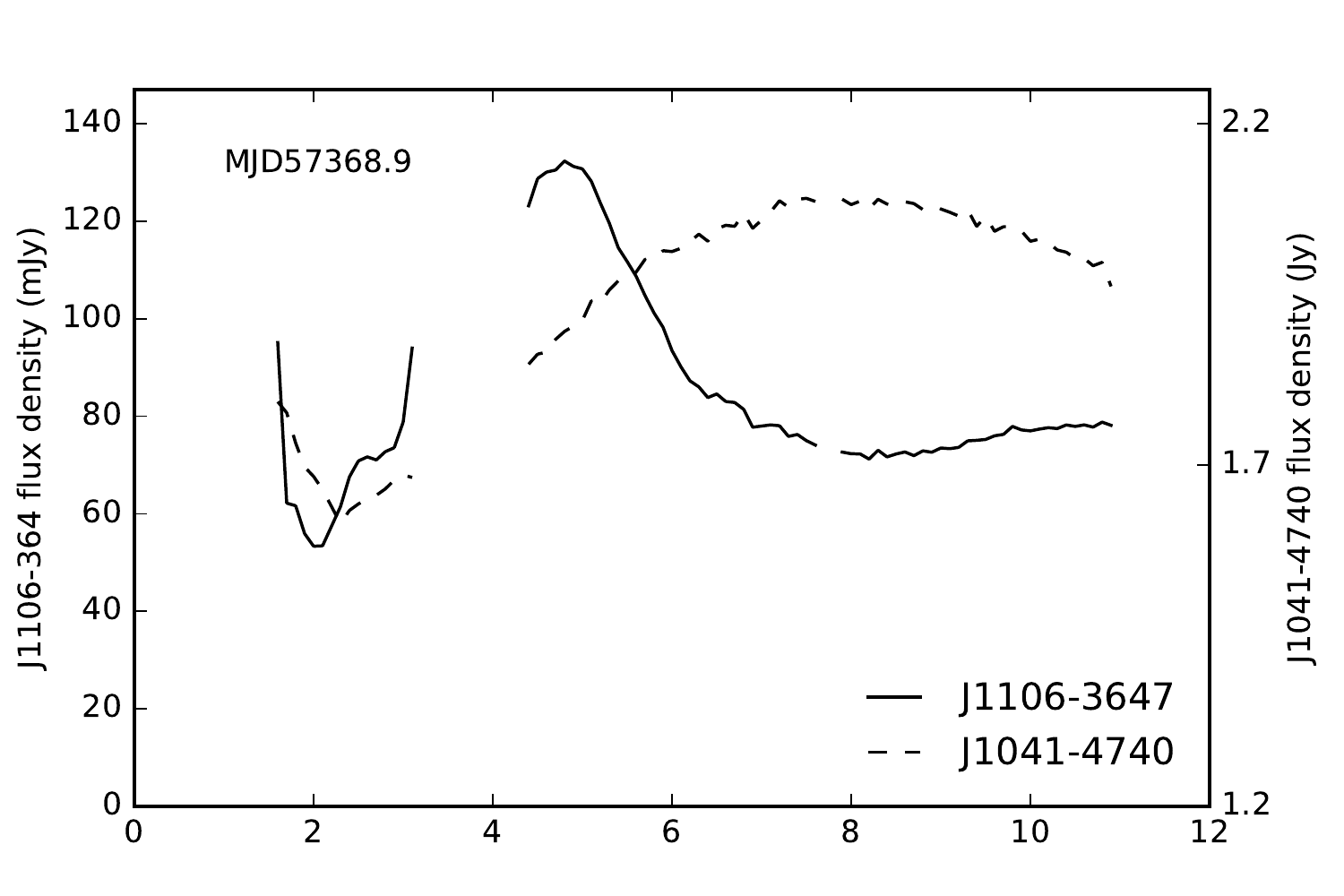}
\caption{Four examples of \J1106 spectra in our dataset {\mods along with the spectra of the control source ICRF J104142.9$-$474006 {\mmods (which is also variable)}; the spectra in each pair are obtained within minutes of each other}. The MJD of each spectrum is indicated on the top right of each {\mods plot}. Small gaps in the spectra are where radio-frequency interference was flagged by our processing pipeline, and the affected data were excised. The thermal noise level is below the width of the lines.}
\end{figure*}

\subsection{Statistics of the variability}
As we have a large number of ATCA spectra, we can evaluate the mean and root-mean-square (r.m.s.) of the \J1106 flux-density at a wide range of radio-frequencies; these estimators are shown in Fig.~2. The mean spectrum is, by virtue of our source selection criteria, fairly flat in our survey band (4-8~GHz). And below 4~GHz it declines gently. By contrast, the r.m.s. spectrum has a clear peak around 5~GHz, where its value is approximately 35\% of the mean flux-density. Towards higher frequencies, the r.m.s. declines steadily, reaching 15\% of the mean flux-density at our highest observing frequency. Below 5~GHz the r.m.s. declines rapidly down to the lower-limit of the ATCA high-frequency receivers, around 4~GHz.

At frequencies of 4~GHz and above, essentially all of the r.m.s. variation shown in Fig.~2 is attributable to the \J1106 radio signals as they arrive at Earth. Familiar systematic measurement errors (calibration errors; pointing errors; atmospheric opacity variations etc.) are expected to contribute, but only at a level $\la1$\% of the mean flux-density \citep{lovelletal2003}. There is, in addition, a systematic error specific to ATCA, described in \S2.2, but that is also at the $\sim1$\% level. 

The gap between 3 and 4~GHz, visible in Fig.~2, is unavoidable, as no ATCA receivers cover that region. Below 3~GHz, there are two main contributions to the r.m.s. flux-density: variations in the \J1106 signal; and source confusion. At low frequencies, our data are susceptible to source confusion because ATCA has poor snapshot imaging capability. (ATCA has only six antennas, and the telescope is usually configured as an East-West array.) As noted earlier, the ATESE project targets were selected to avoid significant confusion in the 4-8~GHz band, but at lower frequencies the primary beam of the telescope increases in size and confusion is much more of a problem. Near 1.4~GHz, there is $\sim100\,{\rm mJy}$ of confusing flux in the primary beam of ATCA, for the field centred on \J1106. We therefore interpret the 1-to-3~GHz r.m.s. as the sum of a rapidly-declining contribution from confusing sources, and a rapidly increasing contribution from \J1106. It is difficult to disentangle these two contributions, with our data, and therefore we restrict our analysis of the \J1106 variations to the 4-11~GHz region --- see \S3.

\subsection{Example spectra}
As noted earlier, our ATCA spectra of \J1106 show a great variety of shapes, and many have features that are highly unusual for AGN. It therefore seems appropriate to give readers a small sample of some of the spectral oddities that \J1106 has exhibited. Those who are interested in viewing more spectra can browse the full dataset --- see Bannister et al (2017, in preparation), for details.

Fig.~3 shows four spectra drawn from the ATESE data on \J1106, on the dates indicated in each panel. These spectra cannot be characterised in any simple way; they all contain a variety of features. In particular we see: broad spectral peaks and troughs (several GHz wide), notably at MJD57057.7 and MJD57368.9; large-amplitude wiggles on a scale $\sim1$~GHz, strongest at MJD57214.4; and abrupt changes in gradient -- kinks in the spectrum -- the most striking of which is seen at approximately 5~GHz, on MJD 57087.6. The genesis of these features, in the context of interstellar scintillation, is discussed in \S3.

\begin{figure}
\includegraphics[width=\linewidth]{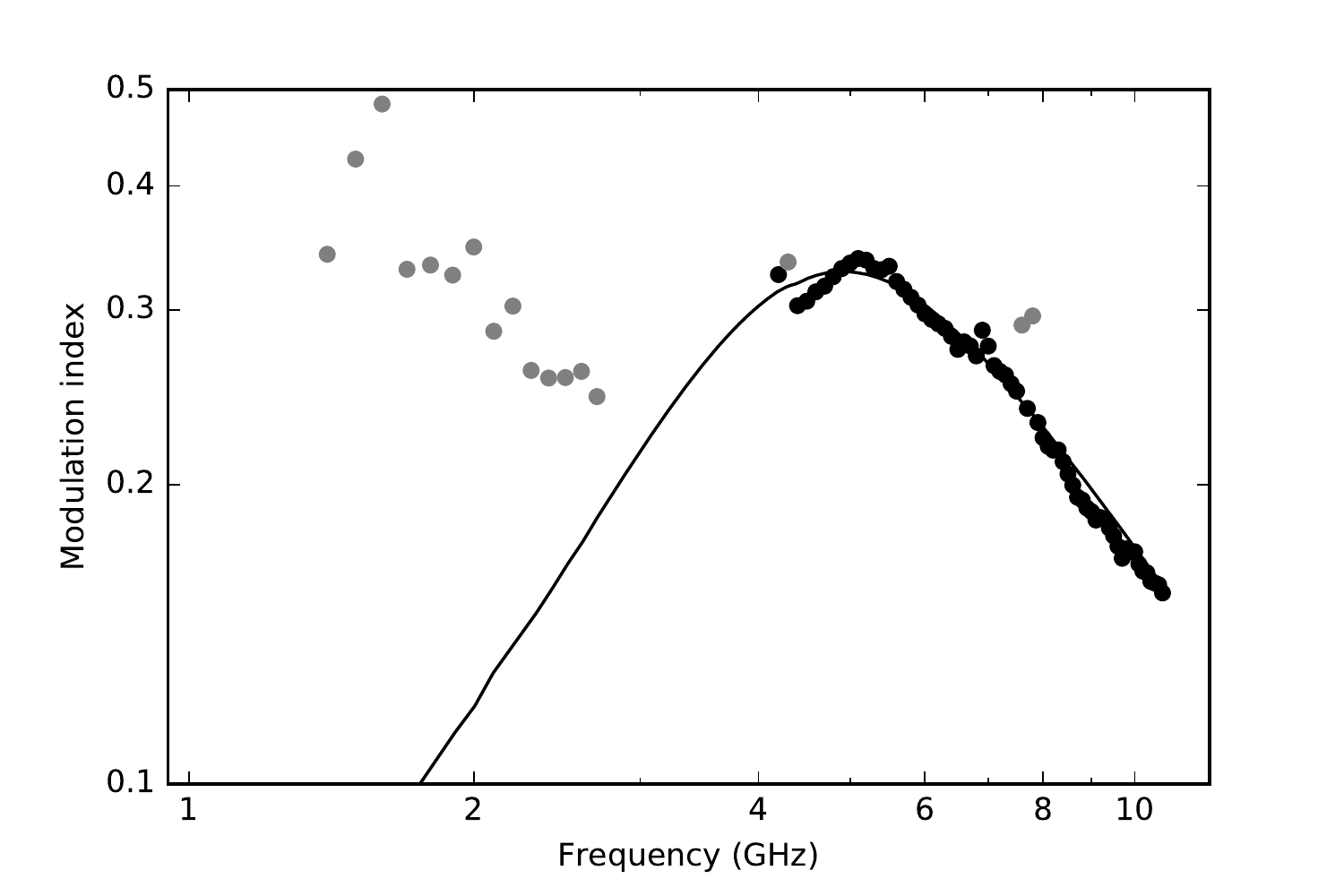}
\includegraphics[width=\linewidth]{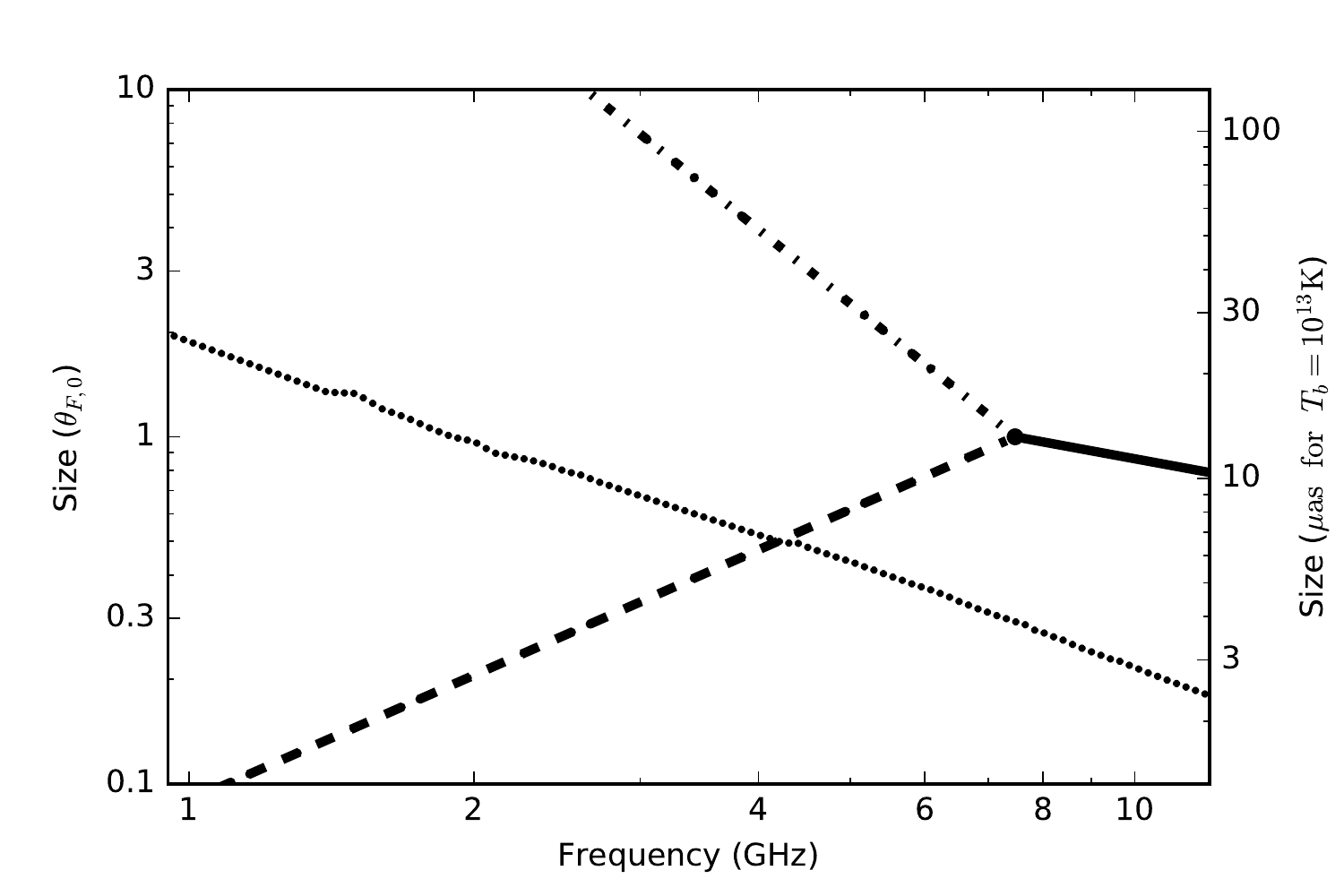}
\caption{The variations of \J1106 compared to a model from GN06. The top panel shows modulation index (r.m.s. over mean), as a function of frequency. The mean and r.m.s. of the \J1106 data used here are as shown in Fig.~2. The model is that of GN06, with {\mods both the compact (i.e. scintillating) fraction and brightness temperature of the compact component independent of frequency. Fitting to the data points shown in black results in the compact fraction of 31 per cent, the transition frequency of $7.5\,\mathrm{GHz}$ and the source-to-Fresnel size ratio of $0.29$ at the transition frequency. Scattering in this model is somewhat weaker than predicted by \citet{walker1998}, whose model suggests a transition frequency $\approx14\,\mathrm{GHz}$ at this location.} The lower panel shows the various angular sizes which are relevant: that of the source itself (dots); the Fresnel scale (solid line), plotted only above the transition frequency; and the diffractive (dashed line), and refractive (dot-dashed line) scales, plotted only below the transition frequency. {\mods As only the ratio of the source size and Fresnel scale is constrained, the size is given in units of the latter at the transition frequency; for $T_b=10^{13}\,\mathrm{K}$ this corresponds to $\theta_{F,0}\approx13.2\,\mu\mathrm{as}$ (cf. Section 5) as shown on the right axis.}}
\end{figure}

In addition to the features just mentioned, most spectra exhibit significant structure on scales small compared to 1~GHz. This high frequency structure is not discussed in \S3, because we suspect that it is probably instrumental in origin. High frequency ripples are sometimes seen as an intermittent artifact in the ATESE project spectra of other targets. The single, strongest ripple, that we often see in our ATESE project spectra, has a period near 300~MHz and an amplitude of order a per cent of the source flux-density. The precise origin is unclear -- possibly a parasitic modulation arising from an external system clock -- but it is certainly instrumental in nature, as it has a well-defined period and is seen in the spectra of many ATESE sources. 

\section{Physical interpretation}
As previously noted, the large amplitude and short time-scale of the flux-density variations of \J1106 cannot be explained as intrinsic variations of a synchrotron source at cosmological distances. Rather, the effects must be attributed to refraction and scattering in the ionized interstellar medium, i.e. \J1106 exhibits strong interstellar scintillation (ISS). The large amplitude of \J1106's variations are unusual even for ISS of AGN; but that is not surprising, for the following reason. \J1106 is an order of magnitude fainter than the sources in which ISS has been studied previously \citepalias{heeschen1984, rickettlaziochigo2006}, and thus is expected to be several times smaller in angular diameter (assuming all AGN have a brightness temperature near the synchrotron self-compton limit). So, providing a suitable scattering screen is in the line-of-sight, that means that the spatial patterns of magnification and wave-interference, which the observer moves through, will be subject to much less smoothing, leading to variations with large amplitudes and short time-scales.

\subsection{Statistical model}
Some of the flux-density excursions of \J1106 are comparable to {\mods its} mean flux-density, over a bandwidth of several GHz around 5~GHz. This {\mods suggests that the scintillation is close to or just in} the strong scattering regime, where the wavefront curvature introduced by the ISM is comparable to the Fresnel curvature (i.e. the curvature of a sphere centred on the observer, in the case of a very distant source). We have attempted to quantify this expectation by fitting our \J1106 statistics to the parametrized models of \citet[][GN06, henceforth]{goodmannarayan2006}.

GN06 provided analytic models for the modulation index (r.m.s. flux-density divided by mean flux-density) spectrum. Two parameters are required to uniquely specify the model, they are: the source size (assumed Gaussian in profile) relative to the Fresnel radius, and the strength of the scattering. Unfortunately, because the sub-milliarcsecond scale structure of radio AGN is not well understood, we have no reliable {\it a priori\/} model for the \J1106 source size versus frequency. We therefore tried {\mods optimising the GN06 models using several different parametric representations of the source size, fitting their parameters simultaneously with the scattering strength, as gauged by the transition frequency}. Our representations characterised the compact fraction (i.e. the fraction of the flux density measured by ATCA that is small enough to scintillate), and the size {\mods or brightness temperature} of the compact component as a function of frequency. {\mods Constant values and} power-law variations with frequency were considered for both these aspects, and we found that (i) a good visual match to the data was readily achieved, and (ii) various combinations of parameters were able to produce similarly good fits.

As an example, Fig.~4 shows the best fit we were able to achieve for a model where compact fraction and brightness temperature are both independent of frequency. The model provides a clear interpretation for the structure of the modulation-index spectrum, as follows. At high frequencies the compact component of the source is smaller than the Fresnel scale, and so the modulation index is close to that of a point-source containing 31\% of the total mean flux-density. Near the transition frequency, between strong and weak scattering, the scintillations are fully developed, and the source remains unresolved, so the modulation-index reaches a peak. However, the source-size increases towards lower frequencies, and at approximately 4~GHz it becomes larger than the diffractive scale (``field coherence scale''); leading to a rapid decline in the modulation-index as the diffractive scintillations are quenched by smoothing. The relationship of the various angular sizes for this model is shown in the lower panel of Fig.~4. Note, however, that there are other  models, producing similarly good fits to the modulation-index spectrum, in which the various angular sizes exhibit a different relationship, and the interpretation of the modulation-index spectrum differs accordingly.

\subsection{Broad-band structure}
Broad-band structure is clearly present in our \J1106 data, with broad peaks and/or broad troughs present in many spectra (e.g. Fig.~3, MJD57368.9). Such structure is a generic expectation for compact radio sources that are being significantly (de-)magnified by inhomogenous plasma structures along the line-of-sight. The frequency dependence of the refractive index of ionized gas means that any plasma lens will have a long focal length at high frequencies, and a short focal length at low-frequencies. And for converging lenses, the lens will be over-focused at sufficiently low frequencies, resulting in a demagnification. Thus even a simple lens and the simplest source (i.e. point-like), can introduce broad-band structure into the observed flux. And there are more possibilities to create broad chromatic structure when one allows for more realistic plasma lenses and more realistic sources.

\subsection{Kinks}
Unlike the broad-band structure -- which the ATESE project was designed to take advantage of -- we did not anticipate that our data would contain spectral kinks. But \J1106 exhibits  several examples; the most striking is seen in Fig.~3, MJD57087.6, at approximately 5~GHz. Even when examined with the full spectral resolution of our data, this feature appears to be a discontinuous change in the local spectral index. It is not obvious what the cause is, but in the vicinity of the kink the flux-density is about twice the mean flux-density, suggesting that the source is highly magnified. (Recall that the statistical model, discussed in \S3.1, placed only 31\% of the flux in a compact, scintillating component; so that component would need 4x magnification in order to double the total flux.) We therefore examined the possibility that kink formation might be associated with the presence of a caustic.

Here the basic point is simply that the caustics of a plasma lens move with frequency, so that for a given source, observed at a given epoch, the image multiplicity may increase above/below a particular frequency. The observed spectrum should therefore change at that frequency, possibly leading to a kink. Our analysis of this circumstance is given in brief form in an Appendix; here we quote only the main results.

For reasons given in the Appendix, we are principally concerned with the influence of a fold caustic, for which the analysis is effectively one-dimensional. We therefore need only consider the one-dimensional surface brightness profile of the source. And we note that, in the vicinity of the caustic, there is a linear relationship between frequency offset and position in the source plane, $x$ (measured perpendicular to the fold). We find that a source with a step-function boundary produces a spectral kink, but the kink is too large --- the change in spectral gradient is infinite, whereas \J1106 exhibits a finite jump in the gradient. On the other hand, a source profile that is a linear function of position, at its boundary, yields a smooth spectrum with no jump in gradient at all. Clearly, the requisite source profile is in some sense intermediate between these two cases. We find that a source intensity profile that is locally $\propto\sqrt{x}$ yields a finite step in the spectral gradient, and can therefore explain the \J1106 data. Such a profile is easily realised in practice --- e.g. it can arise as the one-dimensional projection of the curved boundary of a source that has a uniform surface brightness. Importantly, though, the source boundary must be locally concave, in order to obtain the observed sign of the discontinuity in the spectral gradient. Physically, this could perhaps be realised as a sharp, inner edge of a curved jet.

Although this model of spectral kink formation is simple, and appealing, it is somewhat disconnected from the statistical model of the variations described in \S3.1. We note, in particular, that the spectral kink seen in Fig.~3, MJD57087.6, is very close to the location of the peak in the modulation index spectrum, shown in Fig.~4. (And that remains true if if the MJD57087.6 data are excluded from the r.m.s. evaluation.) A caustic interpretation of the kink offers no immediate explanation for the coincidence. 

\subsection{Wiggles}

Wiggles are sometimes seen in our ATCA spectra of \J1106 --- e.g. Fig.~3, MJD57214.4 in the region 4-7~GHz. These spectra are extracted from the ATCA interferometric visibilities under the assumption that \J1106 is the only source in the field, and if there are other (``confusing'') sources present then they could be responsible for the spectral wiggles. We are confident that the wiggles seen in Fig.~3, MJD57214.4, are not due to source confusion, for the following reasons. First, we have imaged the field (on MJD57734) and find a total confusing flux of less than 2~mJy, spread over four sources (individual fluxes: 0.76, 0.53, 0.44, and 0.26~mJy). If these sources are steady, they cannot be responsible for the spectral wiggles evident on MJD57214.4, whose amplitude is $\sim5\,$mJy. We have examined the possibility that any one of these confusing sources might have been as bright as 5~mJy on MJD57214.4, but even in that case the induced spectral wiggles are only a few mJy peak-to-trough, and their shape is quite different to those actually observed. 

Unlike the broad-band structure, and the spectral kinks, we are not aware of a simple interpretation of spectral wiggles in the context of plasma lensing.

One possible interpretation is suggested by examining the statistical model shown in Fig.~4. Below 7~GHz, that model reproduces the \J1106 modulation index spectrum via a combination of refractive and diffractive scintillations. We have already discussed ways in which refractive scintillations -- i.e. focusing and defocusing of the waves -- can affect the spectra. Diffractive scintillations are fundamentally different: they arise from wave interference. Diffractive scintillation is familiar in the context of pulsar dynamic spectra, where it is responsible for the narrow-band structure that is commonly observed. But it is usually not detected in the case of AGN, where the much greater angular size of the source, relative to pulsars, leads to negligible visibility for the interference fringes. In the case of \J1106, Fig.~4 suggests that the source may be smaller than the diffractive scale, and therefore small enough to exhibit interference fringes, at frequencies $\ga4\,$GHz --- at least for the GN06 model shown in the figure. In that model, it is the decrease in fringe visibility that is responsible for the low-frequency decline in the modulation-index spectrum. 

For that same theoretical model, the transition to strong scattering occurs around 7~GHz. Above that frequency we expect there to be only a single image present, and no interference fringes can develop. Asymptotically, the  bandwidth of the fringes is expected to decrease rapidly as one moves to lower frequencies, and one estimates a fraction $\sim10$\% near 4~GHz, for a Kolmogorov spectrum of plasma density fluctuations. This is broadly consistent with the MJD57214.4 spectrum of \J1106 (Fig.~3). {\mods The diffractive timescale, however, does not vary very rapidly with frequency and is expected to be of order hours for a screen distance of $\sim50\,\mathrm{pc}$ (Section 5) and effective transverse velocity $\sim20\,\mathrm{km}/\mathrm{s}$. Thus no detectable trends are expected within our four-minute integrations.} However, we caution that at frequencies just below the strong- to weak-scattering transition, the refractive and diffractive scintillation branches are not well separated, and estimates based on asymptotic results -- valid in the limit of very low frequencies -- should not be taken too seriously.

Overall, wave interference appears to be a plausible interpretation of the spectral wiggles that \J1106 sometimes exhibits.

\section{VLBI Observations}
Because we were confident that the variations of \J1106 should be attributed to refraction and scattering by inhomogeneous plasma, we undertook VLBI observations in an attempt to obtain constraints on the plasma structures from direct imaging. 

We performed phase-referenced observations of \J1106 with the NRAO VLBA over 17 epochs (VLBA project code BB366). We used the digital downcoverter mode which provides 8x32 MHz channels (4 each for RR and LL polarisations) in a single band. Within the constraints of the receiver system, and known bad frequencies, we spaced the channels as widely as possible across the maximum of 384~MHz, in order to best constrain the spectral properties of the image structure. We used 6 tunings, centered on $1446 \pm 96$~MHz, $2346\pm96$~MHz, $4292\pm 192$~MHz, $5320\pm 192$~MHz,  $6342\pm 192$~MHz and $7366 \pm 192$~MHz. We used J1113-3549 as a phase reference and J1108-3550 as a check source. We cycled through each of the tunings with three, one minute observations of \J1106 bracketed by one minute observations of the phase reference, followed by a bracketed observation of the check source. We completed all tunings in approximately 1~hr.  At this declination, and with such short observations, the $uv$-coverage of the baselines is poor.

We reduced the data in a standard manner described in \citet[supplementary]{bannisteretal2016}. Briefly, we calibrated the data using $T_{sys}$ measurements, and standard fringe fitting on the phase reference. Using an interactive script for {\sc AIPS} \citepalias{vanmoorseletal1996} written in {\sc ParselTongue} \citep{kettenisetal2006} we imaged and cleaned the target and applied several rounds of phase-only self-calibration to determine antenna gains. We applied a final round of phase and amplitude self-calibration before producing a final map and fitting a Gaussian component to measure the flux and astrometry. We calculate the systematic errors in the astrometry (which are far larger than the formal errors) based on empirical relations \citepalias{pradelcharlotlestrade2006}.

In all of our observations, the measured visibilities are consistent with a point-like source. Any multiple imaging must therefore be on an angular scale that is small compared to the synthesised beam size (from $3\,\mathrm{mas}$ by $2\,\mathrm{mas}$ at $7.5\mathrm{GHz}$ to $15\,\mathrm{mas}$ by $10\,\mathrm{mas}$ at $1.5\,\mathrm{GHz}$). Nor were we able to detect any astrometric shifts of the unresolved image. Our astrometry is, however, not very accurate, because \J1106 is so far South that the VLBA can only see it at low elevations, with long atmospheric path lengths, and correspondingly large systematic errors introduced by the troposphere  and the ionosphere. At frequencies near $7\,\mathrm{GHz}$, where the combination of tropospheric and ionospheric errors is expected to be minimized, we find that any astrometric shifts in \J1106 must be smaller than $\la5\,\mathrm{mas}$ in right-ascension and $\la10\,\mathrm{mas}$ in declination.

\section{Discussion}
The astrometric limits just quoted translate to physical constraints on the refracting/scattering screen. A useful reference point is the angular Fresnel scale at 7.5~GHz, which is $0.1/\sqrt{D}\,{\rm mas}$, where $D$ is the screen distance in pc. This is the expected magnitude of the image wander, for \J1106, based on the model shown in Fig.~4. To be consistent with our VLBA astrometric constraints then requires the screen to be more than $10^{-4}\,\mathrm{pc}$ distant. This is not very constraining, considering that the closest screens that have been reported for other sources are at distances of a few parsec \citep{bignalletal2006, debruynmacquart2015}.

The particular model shown in Fig.~4 can be used to determine the screen distance, as follows. We assume that the source is emitting synchrotron radiation, at close to the limiting brightness temperature of $10^{12}\,{\rm K}$, and we allow for a modest boost due to relativistic motion towards the observer, increasing the apparent brightness temperature to $T_b\sim10^{13}\,{\rm K}$. The corresponding source radius at 4~GHz is {\mods $3.3\times10^{-11}\,{\rm rad}$}. Fig.~4 (lower panel) shows that this is just half the Fresnel scale at 7.5~GHz, so a screen at roughly {\mods $50\,{\rm pc}$ from us (or, for lower $T_b$, closer)} would be appropriate for the statistical model shown in Fig.~4. This is consistent with our astrometric constraints, given immediately above. 

The interpretations given in \S3, for the spectral variations of \J1106, are sensible working hypotheses. However, we caution that it is possible to fit the modulation index spectrum of \J1106 using a model source that is very different from the one shown in Fig.~4. In particular, one can obtain a similarly good fit to the modulation index spectrum, using the GN06 formulae, with a source that at all frequencies is larger than the diffractive scale. For such a source the interpretation of spectral wiggles given in \S3.4 is inapplicable, because the visibility of the interference fringes would be near zero. It is not clear how to interpret the spectral wiggles if they are not interference fringes, and that is perhaps a reason to favour a smaller source, such as shown in Fig.~4. {\mods  More generally, it would be helpful to obtain independent constraints on the angular size of the source, as a function of frequency, in order to break this degeneracy in the interpretation. Unfortunately, ground-based interferometry cannot resolve the compact source components which are the strongest scintillators.}

Whatever the physical origin of the spectral wiggles, the fact that they are commonly present in \J1106 spectra suggests an alternative interpretation for the spectral kinks (e.g. Fig.~3, MJD57087.6). Rather than being associated with a caustic crossing, as we proposed in \S3.3, kinks might simply be cuspy, small-scale peaks, superposed on a smoother, broad-band structure. After all, there is no {\it a priori} reason to suppose that the spectral wiggles should be absent when the source is bright. This point has extra weight for the particular model shown in Fig.~4, where wave-interference is responsible for the wiggles, because the GN06 models describe the effects of a Kolmogorov spectrum of plasma inhomogeneities, for which caustics are not expected to arise.

It is worth noting that the GN06 models describe an isotropic, Kolmogorov spectrum of plasma inhomogeneities. By contrast, the well-studied IDV sources, and the pulsar parabolic arcs, both point to highly anisotropic screens {\mods (of $\sim10^{1\pm1}\,$AU dimensions)}. It would therefore be of interest to determine a set of approximations, like the GN06 formulae, that are appropriate to the case of one-dimensional screens.

Following the recognition that diffractive scintillation might be playing a role in the observed spectra of \J1106, we undertook an eight-hour synthesis with ATCA on 12th December 2016, covering the 4-11~GHz region. Unfortunately this observation revealed no significant variations of any kind in \J1106. It seems that this source has now stopped scintillating. Although disappointing, this result was not surprising: a multi-epoch study of a large number of sources has previously demonstrated that IDV is often a transient phenomenon \citep{lovelletal2008}.

\section{Conclusions}
\J1106 is a spectacular scintillator, manifesting large changes in amplitude and spectral shape on short time-scales.  Our ATCA spectra are often rich in structure, and are unlike any other radio spectra that we are aware of for an AGN. However, to date there have been few detailed studies of scintillators with wide-band radio receivers such as those now available on the ATCA; we therefore encourage more studies of this kind, to help elucidate the physical mechanisms responsible for this type of behaviour. There is no doubt that complex spectra can arise from interstellar scintillation, but identifying the precise origin of various features is challenging, particularly near the boundary between strong- and weak-scattering, which seems to be relevant to our \J1106 spectra. We propose that wave-interference is responsible for some of the observed spectral structure, implying that the source is small compared to the field-coherence scale, at some frequencies. However, \J1106 is relatively faint, and our data are consistent with a mildly boosted synchrotron source (i.e. brightness temperature $\sim10^{13}\,{\rm K}$), seen through a screen at {\mods $\sim50\,{\rm pc}$} from us.

\section*{Acknowledgements}
{\mods We thank the referee for detailed comments which helped improve the manuscript. }The Australia Telescope Compact Array is part of the Australia Telescope National Facility which is funded by the Commonwealth of Australia for operation as a National Facility managed by CSIRO. Observations reported in this paper were made under ATCA project codes C2914 and C2965. This work made use of the Swinburne University of Technology software correlator, developed as part of the Australian Major National Research Facilities Programme. This work was supported by resources provided by the Pawsey Supercomputing Centre with funding from the Australian Government and the Government of Western Australia.

We thank the VLBA director for granting Director's Discretionary Time for this
project (BB366). The National Radio Astronomy Observatory is a facility of
the National Science Foundation operated under cooperative agreement by Associated
Universities, Inc.



\bibliographystyle{mnras}




\appendix
\onecolumn
\section{Spectral kinks from caustic crossings}

\noindent The kink in the spectrum of \J1106 observed on MJD57087.6 can potentially be explained {\mods refractively} as a caustic crossing. Caustics are the loci of points in the source plane where the lens mapping is degenerate and the magnification formally diverges. The location of a caustic with respect to the source depends on the wavelength, which translates spatial variation in the magnification factor into structure in the instantaneous spectrum. Many types of caustics are possible. 
 In the present context, the most important are the lowest order catastrophes -- folds -- because they form sheets within the three-dimensional space of source-plane position and observing frequency. In that space, cusps form lines, and higher order catastrophes are confined to points.  

In the source plane, a fold caustic can be locally approximated as a one-dimensional structure. The magnification across the caustic curve diverges, and when a point source crosses the line a pair of infinitely bright images is created (or destroyed, depending on the direction of crossing). This divergence is moderated by the finite brightness temperature of any real source, as an infinite magnification resolves arbitrarily compact sources. On the contrary, the magnification in the direction along the fold caustic remains finite (when away from cusps),  and its other optical properties vary slowly along the curve. Therefore, the total observed flux density is given by convolving the intensity, $I_\nu(x,y)$, projected along lines $x=\const$ parallel to the fold
\bea
G_\nu(x)=\int\md y\, I_\nu(x,y),
\eea
with the magnification factor $\mu(x)$:
\bea\label{Fnu}
F_\nu(x_0)=\int\limits_{x>x_0}\md x \,\mu(x-x_0)G_\nu(x),
\eea
where $x_0$ is the location of the fold. The magnification divergence is universal for all folds {\mods \citep[e.g.,][]{arnold1984}},
\bea
\mu(x)=\sqrt{\frac{2\rho}{x}},\, x>0,
\eea
where the characteristic scale of the fold $\rho$ is given by the inverse second derivative of the deflection angle at the location of the caustic image, $\rho=1/|\alpha''|$ {\mods \citepalias{schneiderehlersfalco1992}}. There is also another, regular image present for both $x>0$ and $x\leq0$ and therefore contributed by portions of the source on either side of the caustic, but its brightness varies smoothly across the fold and is not of interest to us.

\begin{figure}
\includegraphics[width=0.48\linewidth]{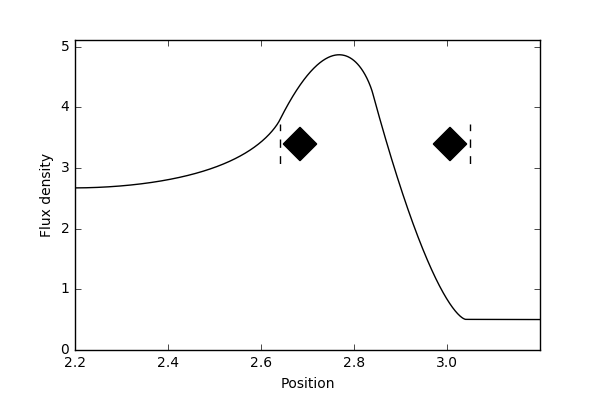}
\includegraphics[width=0.48\linewidth]{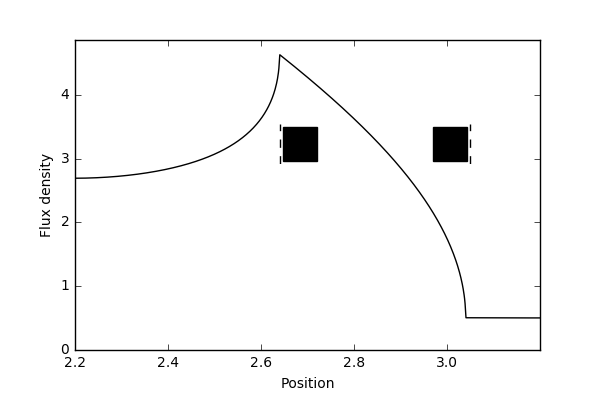}\\
\includegraphics[width=0.48\linewidth]{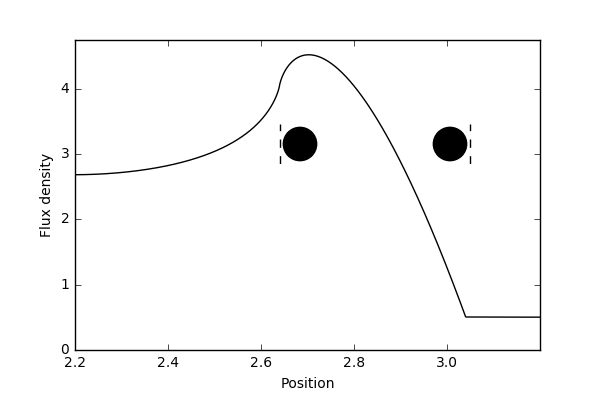}
\includegraphics[width=0.48\linewidth]{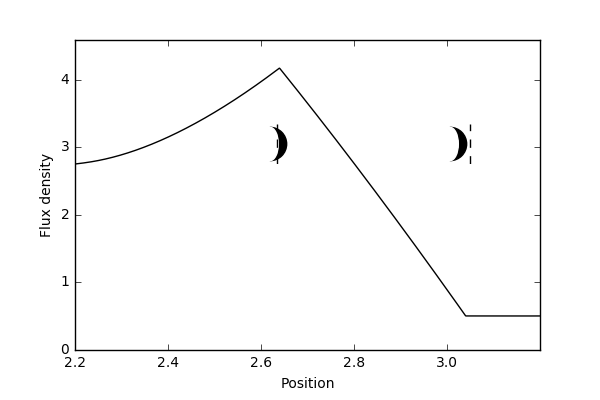}
\caption{The total flux density of differently shaped sources, as marked, in transit of a fold caustic. The mutual orientation of the source and the caustic is as shown in the insets, where the region with extra images is to the right of the dashed line (relevant for the bottom right plot). The plots illustrate that: (i) a sharp boundary is necessary but not sufficient for a discontinuity of the derivative (`a kink') at contact with the caustic; (ii) for the discontinuity to be finite the boundary has to be locally smooth at the contact point and (iii) only a locally concave boundary can induce a finite decrease in the derivative.}\label{figure:convulosions}
\end{figure}

The divergence of the magnification is mild enough that the convolution~(\ref{Fnu}) remains continuous at $x\to x_0$ for any finite $G_\nu(x)$. This is not necessarily the case for the derivative though, as Fig.~\ref{figure:convulosions} illustrates for various projected intensity profiles. It can be shown that a finite discontinuity in $G_\nu(x)$ (which is a projection, so imagine a rectangular source, with one side exactly parallel to the fold, {\mods cf. Fig.~\ref{figure:convulosions} top right}) yields ingress and egress terms in the light curve that are proportional to $\pm\sqrt{x_\pm-x_0}$, where $x_\pm$ are the positions of source edges parallel to the caustic. The derivative of the square root diverges at the zero of its argument, and therefore although a kink forms in this case, the jump in the derivative is infinite at the kink, which is at odds with the appearance of the \J1106 data.

A less restrictive, though hardly more natural, model of the source is a rectangle whose boundaries are not aligned with the caustic, so that $G_\nu(x)$ increases linearly from the point of first contact {(\mods cf. Fig.~\ref{figure:convulosions} top left)}. Such projected brightness profiles result in ingress/egress terms proportional to $\pm(x_\pm-x)^{3/2}$, which are smooth across the fold line and thus do not form kinks at all. By the same token, any source with a projected surface brightness that is linear in $x$ will not yield a spectral kink as it crosses a fold; this rules out a large class of source structures.

Interestingly, a source of constant surface brightness with a smooth, curved boundary {(\mods cf. Fig.~\ref{figure:convulosions} bottom left)} is just right to give a finite kink --- that is, a finite discontinuity in the gradient of the flux-density versus frequency. For such sources the projected intensity near the contact point is 
\bea\label{Gsmooth}
G_\nu(x)\simeq I_\nu2\sqrt{2Bx},
\eea
where $B$ is the curvature radius of the boundary. The flux density in the newly created images near the ingress is then
\bea\label{Fjump}
F_\nu(\Delta x)\simeq\left[\begin{array}{ll}2\pi I_\nu\sqrt{B\rho}\Delta x & \Delta x>0,\cr 0 & \Delta x\leq 0\end{array} \right.
\eea
where $\Delta x$ is the depth to which the source edge is inside the extra images region, which does have a finite jump of the derivative at the derivative crossing. This is \emph{not} the kink we are after though as this finite discontinuity in the derivative appears at a local minimum and not local maximum as seen in the data -- in other words, it is a drop in the derivative that we see in the data rather than a jump as~(\ref{Fjump}) suggests. The final contact, the end of the egress, which one might have expected to be close to a local maximum, is not helping either due to asymmetry of the two sides of the caustic. 
{\mods It can be shown that the flux density 
 is smooth on the final contact; 
its derivative diverges logarithmically as the final contact is approached but as this position 
is also an inflection point of $F_\nu(x)$ the curve just grazes its vertical asymptote and continues on the other side.}

To get a drop in the first derivative rather than a jump, one needs to invert the sign of~(\ref{Fjump}), which is possible by assuming that we are looking at a hole, or a shadow with a smoothly curved, but sharp boundary {(\mods cf. Fig.~\ref{figure:convulosions} bottom right)} . That is equivalent to a negative-brightness source superimposed on a bigger, positive intensity one. Another way of saying this is that the relevant source boundary must be locally concave. Such boundaries occur in crescent-shaped sources, for example; {\mods indeed, similar signatures can be found in simulated light curves of gravitationally microlensed planetary crescents \citep[see Fig.~4 of ][]{ashtonlewis2001}}. 
The magnitude of the drop in the derivative depends on the curvature radius and the brightness temperature difference between the dark inside and bright outside of the boundary via~(\ref{Fjump}), and could potentially be inferred from the measurement of the magnitude of the derivative drop if there is an independent way to constrain the characteristic scale of the fold $\rho$.

{\mods
As the observable flux density~(\ref{Fnu}) depends only on the projection of the source surface brightness along the fold line (or, more precisely, along the $\mu=\const$ lines) the case of a concave source boundary crossing a straight caustic is formally equivalent to that of a caustic curved away from a source with a straight boundary (or even a convex one, just less curved than the caustic). This curved fold would need to have its extra images region on its concave side, which is possible in plasma lensing. However, with curvature much greater than that of the source edge, such a caustic would necessarily affect only a small area of the source. Therefore, although able to produce small peaks in the spectrum, this model can hardly account for the large-scale kink observed in \J1106 when it is highly magnified.}

\bsp	
\label{lastpage}
\end{document}